\documentclass[manuscript,screen]{acmart}
\AtBeginDocument{%
  \providecommand\BibTeX{{%
    \normalfont B\kern-0.5em{\scshape i\kern-0.25em b}\kern-0.8em\TeX}}}

\setcopyright{acmcopyright}
\copyrightyear{2023}
\acmYear{2023}
\acmDOI{XXXXXXX.XXXXXXX}

\acmConference[CHI ’23]{Make sure to enter the correct
  conference title from your rights confirmation emai}{April 28,
  2023}{Hamburg, Germany}
\acmBooktitle{Supporting Social Movements through HCI and Design Research Workshop at CHI Conference on Human Factors in
Computing Systems (CHI ’23),
April 28, 2023, Hamburg, Germany} 
\acmPrice{15.00}
\acmISBN{978-1-4503-XXXX-X/18/06}

\begin{document}

\title{Young Humans Make Change, Young Users Click: Creating Youth-Centered
Networked Social Movements}

\author{Mina Rezaei}
\affiliation{%
  \institution{Department of Human Ecology, University of California, Davis}
  \streetaddress{1 Th{\o}rv{\"a}ld Circle}
  \city{Davis}
  \country{USA}}
\email{minrezaei@ucdavis.edu}
\author{Patsy Eubanks Owens}
\affiliation{%
  \institution{Department of Human Ecology, University of California, Davis}
  \streetaddress{1 Th{\o}rv{\"a}ld Circle}
  \city{Hekla}
  \country{USA}}
\email{peowens@ucdavis.org}


\renewcommand{\shortauthors}{Rezaei and Owens.}

\begin{abstract}
  
From the urbanists’ perspective, the everyday experience of young people, as an underrepresented group in the design of public
spaces, includes tactics they use to challenge the strategies which rule over urban spaces. In this regard, youth-led social movements
are a set of collective tactics which groups of young people use to resist power structures. Social informational streams have
revolutionized the way youth organize and mobilize for social movements throughout the world, especially in urban areas. However,
just like public spaces, these algorithm-based platforms have been developed with a great power imbalance between the developers
and users which results in the creation of non-inclusive social informational streams for young activists. Social activism grows agency
and confidence in youth which is critical to their development. This paper employs a youth-centric lens—which is used in designing
public spaces—for designing algorithmic spaces that can improve bottom-up youth-led movements. By reviewing the structure of
these spaces and how young people interact with these structures in the different cultural contexts of Iran and the US, we propose a
humanistic approach to designing social informational streams which can enhance youth activism.
 
\end{abstract}





\maketitle

\section{Introduction}

Social movements have traditionally defined themselves as occupying public space \cite{Castells15},\cite{Berman83}. Youth-led revolutions and social movements around the world are associated with specific places, such as Tahrir Square and the Egyptian revolution (2011) and Wall Street in the Occupy movement (2011). Urbanists have highlighted the role of public spaces in youth development. They assert that youth use of space is distinct from other groups’ use of space and is a form of resistance against institutionalized power. For instance, Certeau  (1984)\cite{Certeau84} employs "strategies" to describe the hegemonic power structures that discipline public realms and "tactics" for the youth's way of being in a public space. Urbanists\cite{Cox20},\cite{Rezaei15} propose a "bottom-up" approach to creating inclusive urban spaces for young people. They suggest that for a youth-friendly urban space, designers should recognize the younger generation's spatial preferences and tactics by including youth in the design of the space.

With the advancement of ICT, according to \cite{Castells15}, \cite{Castells09}, social movements have been characterized by a decentralized networked form of organization that allows for increased flexibility and effectiveness in mobilizing people for collective action. These networked organizations allow individuals to claim power and challenge dominant forms of authority in a less controlled space. Although there is a less visible strategy in these networked spaces, similar to youth presence in urban spaces, adolescents on social informational streams employ certain tactics to define their collective identity and support their cause. Youth represent their social and political perspectives by posting, sharing, and other social media interactions. However, some research on youth representation on social informational streams suggests that although youth have more digital competency compared to adults, they are not aware of the algorithmic behavior \cite{Boyd16}. On the other hand, there is a social distance between the developers of these platforms and young people as users \cite{Johnson:2012}. Developers are usually unfamiliar with youth tactics of self-representation or their cultural and social background. They consider all youth as "users" of the space without understanding their differences. In this paper, we argue that including youth in the development process of social media platforms and empowering them can foster youth engagement and raise their social and political activity in social informational streams.

\section{Insights}
Taking up space, either to protest or for other forms of representation, is a process of defining youth. \cite{Massey05}. Young people have always been at the forefront of shaping social movements, as it has been fruitful to their self-confidence, self-respect, and other developmental characteristics\cite{Kallio:2016}. In recent years, youth as so-called digital natives\cite{Prensky10} have engaged in political and social issues through social informational streams. Social streams have facilitated youth political agency by providing a faster and easier way to mobilize online and offline campaigns and to reach a larger audience free of charge or with less money \cite{Tarafdar17}. Young people exercise agency, challenge social and political structures, and produce new values in a new public space which is a networked space \cite{Castells15}. While regulations and codes of public spaces are visible and tangible to people, understanding the structure of algorithmic space is challenging. Youth feel freer to share their political and social perspectives in these autonomous spaces \cite{Castells15}. However, social media platforms—like urban spaces—are designed with a top-down approach, which does not prioritize or empower youth activists. Young people use social media with different tactics. Based on their cultural and social background, they go beyond the features designed by the platform \cite{Cornet:2017:PUC:1358628.1358946} creatively to represent their political views. As Castells (2015)\cite{Castells15} puts it, these networked spaces are more spaces of “outrage” by non-transparent algorithmic rules which disrupt the progression of the \cite{Cornet:2017:PUC:1358628.1358946}the movement by distracting and confusing youth or lacking features which would benefit youth activists to broadcast content about their movement. A dialogue between the developers of social informational streams and young people is necessary to create a more youth-inclusive or friendly online social activism experience.
\subsection{Youth tactics and the cultural contexts}
Even though the networked space has its own algorithmic rules and strategies, it can also be controlled by governments. For instance, there has been news about banning TikTok in the US for security reasons\cite{Berman-2023} while it is one of the most popular social media platforms among teens \cite{Pew-Research-Center22}. In some countries, such as Iran, the authorities prohibit social media as they say it threatens social security. However, since these platforms have been banned during different social movements, youth believe social media is blocked because it enables freedom of expression and the opportunity to organize social uprisings in their cities. Using VPNs, young people bypass these restrictions and still share political expressions on blocked social media sites. However, as \cite{Boyd16} says, structures of social segregation, race, and class will be reproduced in online networks. Access to the blocked network is not an option for many youths in disadvantaged communities in Iran—either because of a lack of technical knowledge \cite{Boyd16} of how to use proxies and social media or because of financial problems, since VPNs are not free. The representation of youth from marginalized communities on social informational streams is less pronounced than other groups of young people. As such, social media may not be easily accessible to all people around the world. 

\subsection{Youth use of multimedia to express their social and political opinions }
Youth often innovate social movement media practices \cite{Costanza12},\cite{Cornet:2017:PUC:1358628.1358946}. They use their tactics to find people with similar beliefs and values to build their collective identity\cite{Bernstein05}, \cite{Kallio:2016}. For instance, sharing a clip of George Floyd's tragic death scene sparked large demonstrations against systemic racism in the US and around the world. Or in Iran, during the Mahsa Amini movement—a youth-led movement started in 2022, which seeks to transform dominant cultural and social patterns—a young person made a song based on the tweets of people who shared why they wanted a change on Twitter and Instagram. The tweets started with the hashtag For. The song creator named it “For” and shared it on his Instagram page. It soon became viral through social media and became an anthem for the movement. It helped sustain the movement and construct a collective identity among the young protesters. Social media designers need to explore the different forms of media that youth use around the world and make their platforms more compatible with youth tactics and preferences.

\subsection{Youth use of hashtag to support a social movement}

Hashtags are one of the features available on many social media platforms. They are keywords that assign information to categories \cite{Stefania15} to increase the visibility of topics, connect like-minded people, and initiate a discussion between people with different perspectives around the same issue \cite{Literat19}. Hashtags function to link related information around a specific topic. Most networked social movements are recognized by at least one hashtag. Although users create hashtags, they can be one of the ways social media manipulates social movements \cite{Stefania15}, \cite{Literat19}. Platforms can choose to limit searches by hashtags. For instance, certain hashtags that are considered vulgar are hidden from searches \cite{Literat19}. In both the Black Lives Matter\cite{Pruitt-Bonne-2021}  and Mahsa Amini movements \cite{Kerdo-2022}, users' posts were censored, without considering cultural contexts, because they were considered "sensitive content"\cite{Facebook}. On the other side, in the BLM movement, a group of young people shared images of black squares in solidarity with black victims of police violence on social media. However, they shared these images with BLM hashtags. Consequently, when people searched BLM hashtags, instead of seeing the information about the protest locations, donations, and police brutality documentation, they saw black squares \cite{Vincent-2020}. In the same vein, youth in the Mahsa Amini movement in Iran used Mahsa Amini and other related hashtags in their posts about daily life or other non-related issues and intentionally or unintentionally attached themselves to a popular trend, thus gaining attention and voice \cite{Literat19} while obscuring a channel of information about the movement. Youth need to be aware of how the algorithms of hashtags work, so they can control how to use them more effectively in social movements. Programmers of the hashtags can also apply new changes to the behavior of hashtags based on youth interaction.
\subsection{Social media strategies to direct social movements}
Social media algorithms try to maintain youth outrage during social movements \cite{Castells15} without necessarily helping them to realize their aim. Recommendations, sorting, filtering, ranking functions, and disconnective functions like blocking can facilitate filter bubbles \cite{Cao21}. Trapping youth in like-minded circles that echo their voice and confirm their pre-existing assumptions can lead to radicalization \cite{Conti:2009:DDS:1555009.1555162},\cite{Etter21},\cite{Cao21},\cite{Riebe:2018:} and isolation from reality. In both the BLM (2020) and Mahsa Amini (2022) movements, posts like "Silence is violence", "Silence is supporting injustice", and more radical posts such as "If you are silent, block me", and "Silence or neutrality is equal to cruelty" were shared by many "like-minded" people on different social media platforms. Silence can be due to several reasons, but forcing people to post about a cause or blaming them for being silent is not beneficial to the social movement. This can also lead to the emergence of users who do not believe in a specific social cause. However, they support it out of fear that their friends or followers will ignore them. Young people seek radical societal change, but they are unaware that algorithms can quickly produce fake change. As Beckerman (2022)\cite{Beckerman22} postulates: "Radical change does not start with yelling. It starts with deliberation, a tempo that increases, a volume set first at whispers'' \cite{Beckerman22}. Moreover, the algorithms that control social movements can lessen youths' hope \cite{Castells15}, {\cite{Jenkins-2019} of a significant societal shift. By limiting users' feeds to like-minded posts, social media algorithms degrade their activism to slacktivism which can be easily dismissed by the authorities.  
\subsection{Bots, agents of control in social media }

The other way social media controls youth movements is through bots. Algorithmically controlled social accounts spread disinformation to sway public opinion about a social cause \cite{MarechalNath16}. Also, governments or other power structures can use these bots to distract young activists or suppress movements\cite{stukal22}. Youth can be more affected by bots since they sometimes have problems finding credible sources \cite{Marwik17}. Moreover, emotionally charged information spreads faster\cite{MarechalNath16},\cite{Riebe:2018:}.Young people need to be aware of these mechanisms so they can follow more authentic sources and also be less manipulated by power structures.  

\section{Conclusion and Future Work}
By engaging in social and political causes, young people gain trust and confidence and become more knowledgeable. Social media is one of the main venues for youth to raise social-political issues. They can galvanize youth-led social movements. However, there should be a bottom-up approach to designing these platforms. Designers of these platforms should understand young people's cultural and social differences and their tactics for using social media to support or initiate a social cause. On the other hand, young people should also learn the algorithmic structure of these platforms to make the best use of them for organizing movements. Youth's social and cultural background is one of the main factors in shaping their movements, and this aspect is less represented in the current design of these platforms. Young people also use many creative ways to share their political and social views on social media platforms, which can give design and development ideas to the builders of these platforms. As we discussed, young people's lack of knowledge of the algorithmic strategies of social media platforms can have destructive effects on their movements. The inclusion of young people in the process of making these networked spaces, and empowering them as humans as well as users, could help create more influential social movements that can be recognized in the urban space. Further research is needed to explore young people's use of social media in online and offline social movements in different cultural contexts, their challenges, and the areas to improve the efficacy of social media in organizing the movements. Another area of research can be the disruptive effects of hidden algorithms on youth-led social movements. Further research can also address the feasibility of youth-inclusive social informational streams.

\section{Acknowledgments}
We would like to thank Hau-Chuan Wang, professor of Computer Science at  UC Davis Department of Electrical and Computer Engineering, for his guidance and helpful feedback on this paper.


\bibliographystyle{ACM-Reference-Format}
\bibliography{sample-base}

\appendix

\end{document}